\documentclass[sn-standardnature]{sn-jnl}

\jyear{2021}%

\theoremstyle{thmstyleone}%
\usepackage[version=3]{mhchem} 
\usepackage{amsmath}
\usepackage{bm}
\usepackage{upgreek}
\usepackage{xcolor}
\usepackage{natbib}
\usepackage{float}
\theoremstyle{thmstyletwo}%

\theoremstyle{thmstylethree}%

\begin{document}

\title[Article Title]{Accurate and efficient band-gap predictions for metal halide perovskites at finite temperature}


\author*{\fnm{Haiyuan} \sur{Wang}}\email{haiyuan.wang@epfl.com}
\author{\fnm{Alexey} \sur{Tal}}\email{alexey.tal@vasp.at}
\author{\fnm{Thomas} \sur{Bischoff}}\email{thomas.bischoff90@gmx.de}
\author{\fnm{Patrick} \sur{Gono}}\email{patrick.gono@gmail.com}
\author{\fnm{Alfredo} \sur{Pasquarello}}\email{alfredo.pasquarello@epfl.ch}

\affil{\orgdiv{Chaire de Simulation \`a l'Echelle Atomique (CSEA)}, \orgname{Ecole Polytechnique F\'ed\'erale de Lausanne (EPFL)}, \orgaddress{\city{Lausanne}, \postcode{CH-1015}, \country{Switzerland}}}


\abstract{We develop a computationally efficient scheme to accurately determine 
finite-tem\-pe\-ra\-ture band gaps for metal halide perovskites belonging to the class ABX$_3$ (A = Rb, Cs; B = Ge, Sn, Pb; and X = F, Cl, Br, I).
First, an initial estimate of the band gap is provided for the ideal crystalline structure 
through the use of a range-separated  hybrid functional, in which the parameters are determined 
nonempirically from the electron density and the high-frequency dielectric constant. 
Next, we consider two kinds of band-gap corrections to account for spin-orbit coupling and 
thermal vibrations including zero-point motions.
In particular, the latter effect is accounted for through the special displacement method, 
which consists in using a single distorted configuration obtained from the vibrational 
frequencies and eigenmodes, thereby avoiding lengthy molecular dynamics. 
The sequential consideration of both corrections systematically improves the 
band gaps, reaching a mean absolute error of 0.17 eV with respect to experimental values. 
The computational efficiency of our scheme stems from the fact that only 
a single calculation at the hybrid-functional level is required and that 
it is sufficient to evaluate the corrections at the semilocal level of theory. 
Our scheme is thus convenient for the screening of large databases of metal halide perovskites, including large-size systems.
}

\maketitle

\section{Introduction}\label{sec1}

Metal halide perovskite solar cells have shown remarkable progress in power conversion efficiency, which has been boosted up to 25\% within only a few years \cite{jeong2021pseudo}. Approximately 2000 perovskites \cite{filip2018geometric} can be synthesized but their suitability for photoelectric devices strongly depends on their electronic band gap. In particular, this property varies with the specific phase in which these materials occur, going from 
the highly symmetric cubic phase at high temperatures to structures of lower symmetry at lower temperatures. Given the large variety of available perovskites, it is important to develop a computational tool to screen large databases in search of an optimal material.
Such a tool should provide band gaps at finite temperature in a computationally efficient way without compromising on accuracy. 
In the case of cubic halide perovskites, which have small unit cells, Wiktor \textit{et al.}\  have shown that it is possible to achieve an accurate description of the experimental 
band gaps by combining, on the one hand, high-level many-body perturbation theory for the electronic structure and, on the other hand, {\it ab initio} molecular dynamics to account for the finite temperature, the two schemes implying extensive computational resources of comparable order.\cite{wiktor2017predictive} Next, Bischoff \textit{et al.}\ further demonstrated that the accuracy by which the electronic structure is described could be preserved through the use of computationally more advantageous nonempirical hybrid functionals.\cite{bischoff2019nonempirical} However, this development needs to be accompanied with
a similar reduction of computational cost for treating the effects of finite temperature in order to deploy such methodology to a large set of perovskite materials in an efficient way.

Density functional theory \cite{hohenberg1964inhomogeneous, kohn1965self} based on semilocal approximations for the exchange and correlation energy represents a powerful tool for predicting ground state properties in materials sciences. However, it is well-known that such  semilocal approximations generally underestimate the electronic band gap \cite{sham1983density}. In contrast with this general notion, several previous studies focusing on perovskites have reported that the semilocal functional 
proposed by Perdew, Burke, and Ernzerhof (PBE) gives band gaps in good agreement with experimental data.\cite{even2013importance,yin2014anomalous,demchenko2016optical}
This agreement should be considered accidental because of the neglect of various effects \cite{wiktor2017predictive}, such as spin-orbit coupling (SOC), nuclear quantum effects (NQEs), and thermal vibrations. Many-body perturbation theory in the $GW$ approximation \cite{hedin1965new} is currently recognized as the most accurate scheme for band-gap calculations, especially in its selfconsistent quasiparticle formulation including vertex corrections in the screening \cite{shishkin2007accurate, chen2015accurate, bischoff2021band, tal2021vertex}. However, such calculations are computationally highly demanding and cannot be envisaged as an efficient tool for screening large databases.

Hybrid functionals like the global PBE0 \cite{perdew1996rationale, burke1997adiabatic} or the range-separated HSE \cite{heyd2003hybrid,krukau2006influence} are obtained by admixing a fraction of Fock exchange to the semilocal exchange potential and can serve as valuable alternatives to $GW$ calculations. These hybrid functionals are defined through fixed mixing parameters, and thereby generally fail in providing accurate band gaps for a large class of materials.\cite{he2012screened,chen2012band,franchini2014hybrid}
It has been shown that the higher accuracy can be achieved through material-specific dielectric-dependent 
hybrid (DDH) functionals.\cite{alkauskas2011defect,marques2011density,skone2016nonempirical,chen2018nonempirical,cui2018doubly, zhang2020hybrid, bischoff2021band}
Alkauskas \textit{et al.}\ \cite{alkauskas2011defect} and Marques \textit{et al.}\ \cite{marques2011density} linked the incorporated amount of Fock exchange to the inverse high-frequency dielectric constant $1/\epsilon_{\infty}$. Next, range-separated DDH functionals were developed in which $1/\epsilon_{\infty}$ determines the fraction of Fock exchange in the long range (LR), while retaining a suitable fraction of Fock exchange in the short range (SR).\cite{Abramson2013gap, Abramson2014solid,skone2016nonempirical} Very recently, further developments led to the suggestion of two kinds of nonempirical range-separated DDH functionals.\cite{chen2018nonempirical,cui2018doubly} In the first one (denoted DD-RSH-CAM) the parameters are determined through the first-principles calculation of the dielectric screening function,\cite{chen2018nonempirical} while in the second one (denoted DSH) the parameters are obtained through combining metallic and dielectric screening.\cite{cui2018doubly} 
These two advanced DDH schemes have been shown to yield band gaps of comparable accuracy to 
state-of-the-art $GW$ calculations for a large variety of materials,\cite{chen2018nonempirical,cui2018doubly} but at a significantly lower computational cost.

   

To achieve high accuracy at finite temperature, it is further necessary to account for electron-phonon interactions, which have been found to significantly affect the band gap through the zero-point motions and the thermal vibrations.\cite{wiktor2017predictive} Widely used supercell methods to estimate the electron-phonon interaction are based on the adiabatic approximation and include statistical sampling through Monte-Carlo\cite{monserrat2014electron, patrick2013quantum} and molecular dynamics methods.\cite{pan2014refractive, wiktor2017comprehensive, wiktor2017predictive} 
Within the context of the adiabatic approximation, one also finds the so-called special displacement method (SDM), in which only one single optimal configuration of the atomic positions is sufficient to evaluate the band-gap renormalization due to NQEs and thermal vibrations.\cite{zacharias2016one,zacharias2020theory} 
For polar materials, Poncé {\it et al}.\ pointed out that the adiabatic approximation leads to a divergence at small momentum transfer,\cite{ponce2015temperature} but Zacharias and 
Giustino remarked that the observed divergency is not intrinsic to the adiabatic approximation.\cite{zacharias2020theory} In the SDM, the divergence is found not to build up,\cite{zacharias2016one,karsai2018electron,zacharias2020theory,zacharias2020fully} because the periodic boundary conditions effectively 
short-circuit the long-range electric field associated with longitudinal optical phonons.\cite{zacharias2016one} 
Recent reports show that it is necessary to resort to the nonadiabatic formulation in the evaluation of the zero-point phonon renormalization in the case of infrared-active materials.\cite{ponce2015temperature,miglio2020predominance,shang2021assessment,preprintKresse}
Nonadiabatic effects are usually included through a generalized Fr\"ohlich model,\cite{miglio2020predominance,preprintKresse} which in its one-band formulation can also be used for an estimation of the correction.\cite{nery2016influence,zacharias2020fully,shang2021assessment}
Furthermore, the challenge of calculating band gaps of metal halide perovskites is exacerbated by other issues associated with thermal vibrations, such as anharmonic effects \cite{patrick2015anharmonic}, strong dependence of spin-orbit coupling\cite{wiktor2017predictive} and phonon renormalization \cite{monserrat2016correlation,karsai2018electron,shang2021assessment} on the adopted electronic-structure theory, and cross-coupling between spin-orbit coupling and phonon renormalization.\cite{saidi2016temperature,wiktor2017predictive}

In this work, we achieve accurate band-gap predictions for metal halide perovskites at finite temperature in a
highly efficient way by combining dielectric-dependent hybrid functionals for the electronic structure with the
special displacement method for the nuclear quantum effects and the thermal vibrations.
To highlight the advantage of using dielectric-dependent hybrid functionals, our investigation 
also includes standard hybrid functionals, such as PBE0 and HSE06. 
Our best band-gap predictions are obtained by applying the following procedure. 
We start by calculating initial band-gap estimates with the dielectric dependent hybrid functional 
for the pristine crystalline structure. Next, two corrections are applied. The first correction
results from spin-orbit coupling effects. The second correction accounts for the nuclear quantum effects and the thermal vibrations. In this case, the special displacement method is applied and nonadiabatic effects are estimated through the one-band Fr\"ohlich model.
The band gap of the supercell structure with the special displacements is extracted through a procedure similar to the Tauc analysis of experimental spectra.\cite{tauc1966optical} We demonstrate that both corrections can be determined at the semilocal level of theory, which does not lead to any significant loss of accuracy with respect to hybrid-functional 
results of higher level and thus only implies a moderate computational effort.
The sequential inclusion of the two corrections leads to band gaps that progressively improve with respect to available experimental references, thereby confirming the validity of our approach. 

This manuscript is organized as follows. In Sec. \ref{sec2}, we first describe the model structures and the calculation of the dielectric constants. We then report our band-gap predictions focusing on the corrections due to  
spin-orbit coupling and to electron-phonon interactions. More discussions are drawn in Sec. \ref{sec3}. In Sec. \ref{sec4}, we briefly outline the main theoretical methodologies
used in this work, i.e.\ the construction of the dielectric-dependent hybrid functional and the special displacement method. We also presented the calculation details.

\section{Results}\label{sec2}

\subsection{Structural models}\label{sec2.1}

Our general motivation is to predict band gaps at room temperature for metal halide pervoskites. When various structures of a given material 
exist at room temperature, we take under consideration the available competitive structures to demonstrate the
reliability of our approach upon structural variation. For benchmark purposes, we also consider materials at 
higher temperatures when a detailed experimental characterization is available.

The materials under consideration in this work all have the structural formula ABX$_3$, where A and B are metal atoms and X is a halide atom. 
These materials come in different structures identified by the prefix $\beta$, $\gamma$, $\delta$, R, and M, which indicate that their phase is tetragonal, orthorhombic with corner-sharing octahedra, orthorhombic with edge-sharing octahedra, rhombohedral, and monoclinic, respectively. Within this group, 
all the structures with corner-sharing BX${_6}$ octahedra (R, $\gamma$, and $\beta$) are metal halide perovskites, while the other structures ($\delta$ and M) do not formally belong to this class. Representative atomic structures covering all space groups considered are illustrated in the Supplementary Fig. 1.
 
In this study, we do not consider cubic phases because they are generally unstable at 300 K, 
the targeted temperature for photovoltaic materials. The instability results from soft phonon modes,\cite{saidi2016temperature,huang2014lattice} which can be related to phase transitions.\cite{huang2014lattice}
This indicates that an harmonic description of the phonons would be inappropriate. Indeed, cubic structures are generally subject to important anharmonic effects,\cite{patrick2015anharmonic,cannelli2022atomic} which cause the excess free energy surface to change as a function of temperature.\cite{cannelli2022atomic} The cubic phases at high temperature correspond to average structures, which differ noticeably from the local atomic scale description.\cite{wiktor2017predictive,cannelli2022atomic} For instance, the average B-X-B bond angle corresponds to 180$^\circ$ for the cubic structure, but in molecular dynamics simulations the distribution is peaked at a lower angle ($\sim$168$^\circ$, Ref.\ \citenum{wiktor2017predictive}). 
These structural changes depend on temperature and have noticeable effects on the band gap. Furthermore,
the cross coupling between SOC and thermal effect has been found to be sizable in cubic systems. \cite{saidi2016temperature,wiktor2017predictive} However, our test calculations on noncubic $\delta$-CsPbI$_3$ and R-CsPbF$_3$ show that the error associated to such cross-coupling effects is
$\sim$0.1 eV (see the Supplementary Table 1). For these reasons, cubic structures would require a different theoretical treatment than the one proposed in this work.
 
Here, we make use of the experimentally determined structures given in the Supplementary Table 2.
These structures are used in the calculations of the high-frequency dielectric constant and the band gap. 
Since these structures are taken from experiment, they already include the thermal 
expansion effect corresponding to the temperature at which their structure has been 
characterized (cf.\ discussion in Ref. \citenum{wiktor2017predictive}).

\subsection{Dielectric constants}\label{sec2.2}

The high-frequency dielectric constants used in the DSH functional are calculated
through finite electric fields \cite{umari2002ab}. For fixed atomic positions, we 
calculate the variation of the polarization $\Delta P$ as a function of the electric field 
(cf.\ Supplementary Fig. 2 for $\delta$-\ce{CsPbI3}). More specifically, $\Delta P=P'-P^{\rm{0}}$, 
where $P'$ and $P^{\rm{0}}$ are the polarizations with and without the applied electric field.
The dependence is close to linear up to a critical value, 
above which the calculation no longer converges  \cite{umari2002ab}. 
For all materials, we use in this work a value of 0.001 a.u.\ for the electric field and 
determine the high-frequency dielectric constant through
\begin{equation}\label{Gaussian width}
\epsilon_{\infty} = \frac{4\uppi}{V} \frac{\Delta P}{E_{\epsilon}} + 1,
\end{equation}
where $V$ and $E_{\epsilon}$ are the volume of the supercell and the electric field, respectively. 

The dielectric constants $\epsilon_\infty$ calculated at the PBE level are reported in Table~\ref{DielectricConstant}. 
The values of the dielectric constants used in this work correspond to the average 
over the three principal directions of the dielectric tensor. In these perovskite materials, 
the dielectric constant in a given direction remains generally close to this average value 
(Supplementary Table 3). For instance, the material showing the largest spread is $\delta$-\ce{RbGeI3}, 
for which we found 4.93, 5.81 and 4.75, with an average of 5.16.
Our calculated values show good agreement with previous PBE$+U$ results from Ref. \citenum{petousis2017high}, 
which have been integrated as benchmarks in the Materials Project \cite{jain2013commentary}.
However, there are only few benchmarks for materials belonging to M and $\gamma$ phases.
Therefore, to further validate our calculation of dielectric constants, we take
$\gamma$-\ce{RbGeI3} and M-\ce{CsSnCl3} as representatives of theses phases and 
calculate $\epsilon_{\infty}$  with the higher-level DSH functional that we derived. 
We obtain 5.09 and 3.15, which 
show differences of at most 7\% compared to the plain PBE results of 5.10 and 3.38, respectively. 
This error is noticeably smaller than the typical mean absolute percentage error of 15\% found 
for PBE results compared with experiment.\cite{chen2018nonempirical} Furthermore, the effect of this
error on the band gap is to a large extent attenuated because $\epsilon_\infty$ enters through 
its inverse into the hybrid functional.\cite{chen2018nonempirical}
It has also been shown that the DSH functional with PBE values for $\epsilon_\infty$ is generally 
more accurate for band-gap predictions than a DSH functional with self-consistently 
determined dielectric constants \cite{cui2018doubly,liu2019assessing}. 
Hence, we take $\epsilon_\infty$ values calculated at the PBE level to derive the 
parameter $\alpha_\text{LR}$ of the DSH functionals used in this work.

The inverse screening lengths $\mu$ of these materials vary between 0.67 and 0.82 bohr$^{-1}$, 
with an average value of 0.73 bohr$^{-1}$ (cf.\ Table \ref{DielectricConstant}). This value is close to the average 
result of 0.71 bohr$^{-1}$ that Chen \textit{et al.} found for a large set of materials. \cite{chen2018nonempirical}
To carry out a more detailed comparison, we consider in the Supplementary Table 4 the $\mu$ values for a selection of materials as obtained with three methods, namely Thomas-Fermi, DD-RSH-CAM, and DSH.
We observe that DSH and DD-RSH-CAM give close values of $\mu$ for most materials, which result in similar band-gap estimates. At variance, the Thomas-Fermi values of $\mu$ generally differ more significantly, particularly for wide band-gap materials. These deviations generally lead to inaccuracies as far as the calculated band gaps are concerned. As an example, the reader is referred to the case of LiF discussed in the Supplementary Table 4.

\subsection{Band gaps}\label{sec2.3}

\begin{figure}
    \centering
    \includegraphics[width=0.5\textwidth]{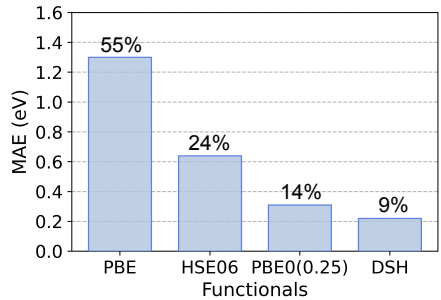}
    \caption{\textbf{Band gap accuracy with respect to ${\bm{GW}}$} Mean absolute errors (MAEs) of band gaps calculated with various functionals with respect to the QS$GW$ values \cite{huang2016electronic,mckechnie2018dynamic}. The percentages correspond to the mean absolute relative errors (MAREs). \label{figure1}}
\end{figure}

The band structures of the class of materials considered in this work, both perovskites and non-perovskites, have been heavily investigated in the literature through a variety of electronic structure methods going from semilocal to $GW$ methods.\cite{huang2013electronic,yu2017oriented,ghaithan2020density,nyayban2020first,su2021stability} The calculated
band gaps typically vary by as much as a factor of two depending on the method.
In our study, we first determine the fundamental band gap of the ternary ABX$_3$ perovskites with PBE, PBE0(0.25), HSE06, and DSH (Table \ref{bare}), including neither spin-orbit effects nor
coupling to phonons. 
As reference, we also list in Table \ref{bare} band gaps calculated through high-level QS$GW$ under the same assumptions.\cite{huang2013electronic, huang2016electronic, mckechnie2018dynamic}
As expected, the functional PBE strongly underestimates the band gaps showing the worst mean absolute error (MAE) of 1.30 eV and the worst mean absolute relative error (MARE) of 55\%. The hybrid functional HSE06 opens the band gap with respect to the PBE value by 0.50 to 1.05 eV, and the hybrid functional PBE0(0.25) further opens the band gap by about 0.7 eV, in accord with general considerations \cite{komsa2010alignment}. 
We remark that the MAEs with respect to the accurate QS$GW$ values progressively decrease 
when adopting in sequence PBE, HSE06, PBE0(0.25), and DSH (cf.\ Table \ref{bare} and Fig. \ref{figure1}). In particular, DSH yields a MAE of 0.22 eV and a MARE of 9\%. 
To further support the quality of the DSH description, we perform for $\beta$-\ce{CsSnI3} 
a comparison between the DSH band structure calculated here and the $GW$ band structure obtained by 
Huang and Lambrecht.\cite{huang2013electronic} As shown in the Supplementary Fig.\ 3, the band gap occurring at the Z point is well reproduced and the overall trends of the band structure are very similar. This level of agreement between hybrid functionals and $GW$ matches that found for other materials in the literature.\cite{deak2017choosing}

We next analyze the general band-gap properties of the materials considered comparing variations of the phase.
As shown in Table \ref{bare}, the R, $\gamma$, and $\beta$ phases all have direct band gaps, while the
$\delta$ phases show an indirect band gap. For the monoclinic phases, the band gaps can be either direct or indirect. 
However, due to the flatness of the bands,\cite{huang2013electronic} the difference between the direct and the indirect bands gaps is generally not large. For instance, in the case of M-\ce{CsSnF3}, we find an indirect band gap of 6.59 eV to be compared with the smallest direct band gap of 6.94 eV. Additionally, we remark that the $\delta$ phases generally have larger band gaps than the corresponding $\beta$ or $\gamma$ phases. This can be readily understood by their band structures. To illustrate this property, we calculate the band structures of $\beta$-\ce{CsSnI3}, $\gamma$-\ce{CsSnI3}, and $\delta$-\ce{CsSnI3} with the PBE functional (cf.\ Supplementary Fig.\ 4). It is found that the valence bands of $\delta$-\ce{CsSnI3} are much flatter bands than those of $\beta$-\ce{CsSnI3} and $\gamma$-\ce{CsSnI3}, which can be related to the distance between nearest neighbor halogen atoms. This provides an explanation for the systematically wider band gaps found for the $\delta$ phase compared to the other phases.

Considering chemical variations within the same symmetry class, we first notice that the calculated band gap $E_{\rm{bare}}$ for all classes decreases for variations of the halide atom going from Cl to I while keeping the composition in terms of metal atoms unmodified. For instance, $E_{\rm{bare}}^{\gamma \text{-}\ce{CsPbCl3}} > E_{\rm{bare}}^{\gamma\text{-}\ce{CsPbBr3}} > E_{\rm{bare}}^{\gamma\text{-}\ce{CsPbI3}}$. This property can be understood invoking the electronegativity, which reduces  from Cl to I. Since the character of the valence band is  dominated by the halide atoms,\cite{huang2013electronic, he2017fully, tao2019absolute} such a downward move in the periodic table results in an upward shift of the valence band and consequently in a narrower band gap. Second, we consider band-gap variations when keeping the A and X composition unmodified and varying the B atom.
In this case, the band gap of Pb-based compounds is always larger than the corresponding Sn-based counterpart. This effect can also be attributed to a reduction of electronegativity going from Pb (2.33) to Sn (1.96). In the absence of any change associated with the halide atom, the smaller electronegativity of Sn leads to an upward shift for both the VBM and CBM.\cite{tao2019absolute} However, Tao \textit{et al.} \cite{tao2019absolute} found that the replacement of Pb with Sn results in a larger shift for the VBM than for the CBM because of their different ratios of $s$ and $p$ character. Therefore, smaller band gaps are found for  Sn-based compounds. Since the electronegativity of Ge (2.01) falls in between those of Pb and Sn, we infer the following trend: $E_{\rm{bare}}^{\delta\text{-}\ce{RbPbI3}} > E_{\rm{bare}}^{\delta\text{-}\ce{RbGeI3}} > E_{\rm{bare}}^{\delta\text{-}\ce{RbSnI3}}$.
Third, when changing the A atom, we don't have a uniform rule for the band-gap variation. Indeed, we find $E_{\rm{bare}}^{\delta \text{-}\ce{RbPbI3}} < E_{\rm{bare}}^{\delta\text{-}\ce{CsPbI3}}$, but we also find $E_{\rm{bare}}^{\delta \text{-}\ce{RbSnI3}} > E_{\rm{bare}}^{\delta \text{-}\ce{CsSnI3}}$. This effect can be related to the fact that the A-atom states lie far away from the VBM and CBM, and
thus do not influence their energy levels directly.\cite{huang2013electronic, he2017fully, tao2019absolute} 

In the following, the calculated band gaps are benchmarked with respect to nine materials for which
a careful experimental characterization of the band gap is available. This experimental set of ABX$_3$ materials comprises eight perovskites and one non-perovskite. The B site is occupied by Ge, Sn, and Pb atoms, while the X site is occupied by Cl, Br, and I atoms. The A site is solely occupied by Cs atoms, but this site is known not to affect to band gap in a significant way.\cite{huang2013electronic, he2017fully, tao2019absolute} The band gaps covered by this experimental set of materials go from 
1.25 to 4.46 eV, which is the relevant range for photovoltaic materials.
Overall, these materials offer a diversified set to benchmark our method.

Many of the materials considered contain heavy elements, which are expected to give significant SOC effects. To estimate SOC corrections, we consider three different functionals: PBE, PBE0, and DSH. The band-gap variations are given in Table \ref{SOC}. We observe similar variations for the three functionals studied. The largest differences between PBE and DSH are found for R-\ce{CsPbF3} and $\gamma$-\ce{CsPbF3} and amount to only 0.10 and 0.09 eV. In a study comprising 103 materials but not including perovskites, Huhn \textit{et al.}\ also found that the SOC corrections calculated with PBE and HSE06 are overall very similar, with most differences falling below 60 meV and the most notable one being 189 meV.\cite{huhn2017one} However, we remark that Wiktor \textit{et al.}\ found that the SOC corrections on the band gap are underestimated by as much as 0.3 eV for Pb-based cubic phases,\cite{wiktor2017predictive} indicating that SOC effects need careful
evaluation for high-symmetry structures containing Pb atoms. Hence, we conclude that the PBE functional is sufficiently accurate to estimate SOC corrections for all considered materials in this work, none of which shows the cubic symmetry.

The calculated SOC corrections can be found in Fig. \ref{figure2}(a) and
Table \ref{final}. These corrections always lead to a band-gap reduction
with respect to $E_\text{bare}$. The largest corrections correspond to Pb-based compounds 
and reach values around $-$1 eV for $\gamma$-\ce{CsPbCl3} and $\gamma$-\ce{CsPbBr3}. The largest correction for a Sn-based compound amounts to $\sim-$0.4 eV and is found for $\gamma$-\ce{CsSnI3}.
We remark that the SOC corrections do not only depend on composition
but also on the underlying atomic structure. 
Indeed, in the case of \ce{CsPbBr3}, we find $-$0.37 and $-$1.02 eV for the $\delta$ and 
$\gamma$ structure, respectively. This indicates that SOC effects may be large and highly nontrivial.

We then obtain the band gaps $E_\text{SOC}$ by adding the SOC corrections to $E_\text{bare}$ (cf.\ Table \ref{final}). 
For the subset of materials for which experimental estimates are 
available, the accuracy of the corrected band gaps improves 
significantly, with the MAE reducing from 0.57 to 0.26 eV. Correspondingly,
the MARE drops from 27\% to 11\%. These errors clearly illustrate the importance of considering SOC effects for achieving accurate band gaps.

In this section, we focus on the coupling to phonons, which affect the band gap through ZPR and thermal vibrations.
For the ZPR and the vibrational properties, we use experimental 
lattice constants at the temperatures given in the Supplementary Table 2. In this way, thermal 
expansion effects are already accounted for. 
For room temperature, we take $T=300$ K in this work. All the modes of the structures considered in this work show positive frequencies.  

To obtain the band gap renormalization caused by zero-point motion and 
thermal vibrations, we use the special displacement method \cite{zacharias2016one} described in Sec. \ref{sec:SDM}.
The supercell sizes adopted in this work contain between 160 and 625 atoms (cf.\ Supplementary Table 5) and have been chosen in order to achieve accuracies better than 0.05 eV. We checked this explicitly for all selected supercells by performing the same calculation with smaller supercells, leading to differences smaller than 0.05 eV for the band-gap predictions. 

The special displacement method does not account for the long-range Fröhlich coupling resulting from the nonadiabatic treatment.\cite{ponce2015temperature,nery2016influence,miglio2020predominance,%
zacharias2020theory,zacharias2020fully,shang2021assessment,preprintKresse} To obtain an estimate for this contribution, we use the one-band Fröhlich model (see Supplementary Note 1).\cite{nery2016influence}  
For a series of perovskites belonging to the class of materials under investigation in our work, we obtain estimates ranging between $-$0.05 and $-$0.13 eV, with a mean 
value of $-$0.09 eV (see Supplementary Table 6). As will be seen in the following, the error resulting from the neglect of this effect is smaller than the final MAE claimed in our 
work (0.17 eV). Hence, it can be concluded that the long-range Fröhlich coupling 
coming from the nonadiabatic treatment of the electron-phonon interaction is 
not susceptible to affect in a significant way the accuracy of our results 
for metal halide perovskite materials.

The SDM method provides supercells with distorted atomic structures, which we use to extract the band gap. More specifically, these disordered supercells simulating the effect of finite temperature cannot be characterised with the same Brillouin zone as the ideal material, because of the lack of translational symmetry. To conform with the practice in the experimental determination of optical band gaps at finite temperature,\cite{tauc1966optical} we therefore rely on extrapolations of the wings in the electronic density of states (EDOS) to obtain the VBM and CBM. For simplicity, we here perform a linear extrapolation of the band wings. \cite{ambrosio2015redox,wiktor2017comprehensive}

In order to perform reliable extrapolations of the band wings, the calculated eigenvalues in the vicinity of the band edges should be sufficiently dense. This condition is not guaranteed for all materials when using the {\sc cp2k} code, which solely uses the $\Gamma$ point. In some cases, the states in vicinity of the band edge are sparsely distributed and do not connect with the band wing upon reasonable Gaussian smearing. It is then necessary to 
use denser {\bf k}-point meshes or alternatively larger supercells.
We illustrate this issue for an ideal structure of $\gamma$-\ce{CsSnI3} in Fig. \ref{figure3}(a). First, we apply the code {\sc vasp} \cite{ kresse1996efficiency, kresse1996efficient} to the primitive unit cell and establish a converged benchmark by increasing the {\bf k}-point sampling of the Brillouin zone (cf.\ Supplementary Fig. 5). Then, we compare the converged reference shown in Fig. \ref{figure3}(a) with a {\sc cp2k} calculation obtained with a 6$\times$4$\times$6 supercell and find good agreement for the band gap within less than 0.005 eV. Hence, the band gap can be obtained accurately provided a sufficiently large supercell is used. In the Supplementary Table 5, we give suitable supercell sizes to obtain converged band gaps for the materials considered in this work. Once a sufficiently large supercell is ensured, the EDOS is determined for the ideal structure and for two displaced structures accounting for ZPR and ZPR+$T$, as illustrated for M-\ce{RbSnF3} in Fig. \ref{figure3}(b). The linear extrapolation of the band 
wings then allows us to extract the VBM [Fig. \ref{figure3}(c)] and the 
CBM [Fig. \ref{figure3}(d)]. Such an analysis is performed for each material (cf.\ Supplementary Figs. 6 and 7).

The use of hybrid functionals for the EDOS with a dense {\bf k}-point mesh or with a large supercell would undermine the efficiency of our band-gap determinations. 
There are reports in the literature with ZPR calculations showing a dependence on the adopted electronic-structure method.\cite{monserrat2016correlation,karsai2018electron,shang2021assessment} Variations are particularly important for silicon \cite{monserrat2016correlation} and diamond \cite{karsai2018electron}
and might amount up to $\sim$0.25 eV. For other materials like MgO and LiF, the variation between functionals 
is much smaller ($\sim$0.05 eV).\cite{shang2021assessment} 
Since it is difficult to develop a rationale for understanding the size of this dependence, we explicitly investigate whether thermal corrections for perovskite materials can be accurately evaluated with the semilocal PBE functional. 
For this, we select M-\ce{RbSnF3} and M-\ce{CsSnF3}, which show noticeable differences between the band gaps 
calculated with PBE0(0.25) and DSH (cf.\ Table \ref{bare}). We check the effect of the functional by extracting 
the ZPR+$T$ band-gap corrections from the EDOS obtained with PBE, PBE0(0.25) and PBE0(0.40) (see Table~\ref{phonon-functional} and Supplementary Fig. 8). The functional PBE0(0.40) can be taken as representative of 
DSH for M-\ce{RbSnF3} and M-\ce{CsSnF3}, since the bare band gaps calculated with these two 
functionals differ by less than 0.08 eV. Additionally, we use PBE and PBE0(0.25) to test R-\ce{CsGeCl3}, 
which shows the largest band gap renormalization due to ZPR+$T$ in our work, i.e.\ of 0.4 eV (cf.\ Table \ref{thermal+ZPR}). In all cases studied, we find that the corrections $\Delta E_{{\rm ZPR}+T}$ obtained 
with PBE coincide with those obtained with the hybrid functionals within at most 0.03 eV.
Thus, this justifies the use of the PBE functional for estimating the effect of the band-gap corrections due 
to ZPR and thermal vibrations in the case of perovskite materials.

The band gap renormalizations due to zero-point motion and thermal vibrations are given in Table~\ref{thermal+ZPR} and are illustrated in Fig. \ref{figure2}(b). 
In particular, we also give the corrections at $T=0$ K corresponding to the ZPR.
The compounds containing the light F atoms show the most prominent ZPR effects, with 
the largest one found for M-\ce{RbSnF3} and amounting to $-$0.20 eV.
The next group of compounds showing sizable ZPRs are those containing Cl atoms, 
among which R-\ce{CsGeCl3} shows the most significant value of $-$0.14 eV.
Increasing the temperature from 0 to 300 K produces further band-gap renormalization.
Together with the ZPR, this leads to $\Delta E_{\rm{ZPR}+\it{T}}$ corrections ranging between $-$0.40 and 0.23 eV (Table~\ref{thermal+ZPR}), which cannot be neglected when aiming at high accuracies. 
We account for the combined effect of SOC and phonons by determining the two band-gap renormalizations 
separately, as these effects have been found to give a negligible cross coupling (see Supplementary Table 1).
Adding $\Delta E_{\rm{ZPR}+\it{T}}$ to $E_\text{SOC}$, we then obtain the band gap estimates 
$E_\text{theory}$, which further improve the comparison with experiment 
leading to a MAE of 0.17 eV and a MARE of 6\% (cf.\ Table \ref{final}). 

The band-gap renormalization due to thermal vibrations can be compared with previous work in the literature. The band-gap variation of $\gamma$-\ce{CsPbBr3} between 0 and 300 K has been investigated through constant-volume molecular dynamics simulations\cite{boziki2021molecular} yielding a shift of 0.01 eV in good agreement with our estimate of $\Delta E_{\it{T}} = 0.02$ eV (cf.\ Table~\ref{thermal+ZPR}). 
Moreover, for large-size nanocrystals of $\gamma$-\ce{CsPbI3}, temperature-dependent luminescence spectra show that the observed band gap remains invariant with temperature.\cite{tang2020high} This behavior stems from the combined effect of lattice expansion and phonon renormalization.
To establish a connection with our work, we estimate the correction due to 
lattice expansion using the experimental thermal lattice-expansion coefficient of $0.39 \times 10^{-4}~\rm K ^{-1}$.\cite{haeger2020thermal} We find a band-gap increase 
of 0.07 eV over a thermal range of 300 K, which almost perfectly compensates the corresponding band-gap reduction of 0.06 eV due to thermal vibrations (cf.\ $\Delta E_{\it{T}}$ in Table ~\ref{thermal+ZPR}). Hence, our correction is fully consistent 
with the null result observed in the experiment.\cite{tang2020high}

Our method relies on the harmonic approximation for describing the vibrational properties of the investigated materials. Patrick et al.\ performed a careful study including anharmonic effects in both 
cubic and noncubic \ce{CsSnI3}.\cite{patrick2015anharmonic} To evaluate possible contributions coming from anharmonic effects, we applied our harmonic scheme to the noncubic $\beta$-\ce{CsSnI3} and $\gamma$-\ce{CsSnI3} at 380 and 300 K, respectively, finding band-gap renormalizations of
0.23 and 0.11 eV due to $\rm{ZPR}+\it{T}$. These values are very close to the respective results 
of 0.16 and 0.11 eV found by Patrick et al.\ including anharmonic effects. 
This suggests that anharmonic effects only lead to small corrections for noncubic perovskites. 
However, we emphasize that anharmonic effects can be larger in the cubic phases because 
the structure then undergoes noticeable distortion moving away from the highly symmetric 
structure, which only holds on average.\cite{cannelli2022atomic}

The correction $\Delta E_{\rm{ZPR}+\it{T}}$ can either increase or reduce the band gap. 
Most materials in our selection exhibit the conventional Varshni effect, 
i.e.\ their band gap decreases with increasing temperature (red shift).\cite{varshni1967temperature} The ZPR and the band gap reduction due to thermal vibrations then contribute in the same direction with corrections being as significant as $-$0.40 eV in the case of R-\ce{CsGeCl3}. 
At variance, some materials, most particularly the Sn-based $\gamma$ and $\beta$ phases, show a band gap opening with increasing temperature (blue shift). This unusual effect is referred to as anomalous band gap shift and has experimentally been observed for a variety of hybrid organic-inorganic lead halide perovskites.\cite{dar2016origin}

In the case of octahedral structures, the origin of the band-gap renormalizations 
can be understood in terms of B-X bond-length and B-X-B bond-angle variations (see Supplementary Table 7). Indeed, it is well-known that reducing the B-X bond lengths or the B-X-B bond angles generally leads to lower band gaps due to enhanced orbital coupling.\cite{coduri2019band, prasanna2017band, knutson2005tuning} 
Practically all octahedral structures in our selection show reduced bond angles and increased bond lengths when going from 0 to 300 K (cf.\ Supplementary Table 7). These trends oppositely impact the band gap with a combined effect that can either lead to band-gap opening or to band-gap closure. More specifically, for the rhombohedral structures, the effect due to the large bond-angle reductions (5$^{\circ}$--10$^{\circ}$) apparently dominates, leading to overall band-gap closing. The larger the bond-angle reduction, the lower the band gap. For $\gamma$ phases, the bond-angle reductions are generally smaller (1$^{\circ}$--4$^{\circ}$) and their effect is similar to the effect due to the bond-length increases, resulting together in small band-gap variations, either red-shifted or blue-shifted.
The structural variations of the $\delta$ phases are generally less significant, with bond-angle reductions and bond-length increases amounting to at most 1$^{\circ}$ and 0.02 \AA, respectively. Nevertheless, their combined effect still yields quite 
sizable band-gap reductions, with the largest being 0.33 eV in the case of $\delta$-\ce{CsPbBr3}.


\section{Discussion}\label{sec3}

In Fig.\ \ref{figure4}, we show how the calculated corrections progressively improve
the theoretical band gaps. We here consider all the materials for which experimental 
band gap values are available. The corrections due to SOC and ZPR$+T$ are added in sequence to the bare band gaps calculated with the DSH hybrid functional. Moreover, our best estimate for the fundamental band gaps $E_\text{theory}$ are compared with optical band gaps from experiment. We hereby neglect excitonic effects,
which are estimated to be on the order of 10 to 100 meV for
this class of materials. \cite{filip2014steric, chen2012photoluminescence, baranowski2020exciton, dias2021role} 
For instance, photoluminescence experiments on $\gamma$-\ce{CsSnI3}
suggest a binding energy of 18 meV.\cite{chen2012photoluminescence}
Calculations based on the Bethe-Salpeter equation also give very small
exciton binding energies for \ce{CsGeX3} (X = Cl, Br, I), on the order of at most $\sim$1 meV.\cite{dias2021role} 

The residual deviations in Fig.\ \ref{figure4} between our theoretical estimates and the experimental values should be assigned to the intrinsic limitations of the adopted theory. The fact that the errors are not systematic 
reinforces this point of view. In fact, neglected effects such as exciton binding energies, polar corrections due 
to LO phonon coupling, anharmonic vibrational effects, errors resulting from the use of PBE for SOC and phonon renormalizations, and cross coupling between SOC and temperature effects, have all been demonstrated to affect the results to a lower extent than the residual error. Residual deviations, such as encountered for R-CsGeBr$_3$, $\gamma$-CsPbCI$_3$, and M-CsSnCl$_3$ should therefore not be attributed to one particular effect. 
It is likely that the largest residual errors ranging up to 0.49 eV for an individual perovskite results 
from limitations of the adopted electronic-structure method. It should be remarked that even state-of-the-art $GW$ calculations show non-systematic behavior with individual errors ranging up to 0.44 eV.\cite{tal2021vertex}

As seen in Fig. \ref{figure4}, the agreement with experiment systematically improves when applying corrections due to $\Delta E_{\rm{SOC}}$ and $\Delta E_{{\rm ZPR}+T}$ in sequence. On average, $\Delta E_{\rm{SOC}}$ and $\Delta E_{{\rm ZPR}+T}$ amount to $-$0.36 and $-$0.15 eV, respectively, corresponding to an average global correction $E_\text{full}$ of 0.51 eV (cf.\ Fig. \ref{figure2}).
The MAEs of $E_{\rm{bare}}$, $E_{\rm{SOC}}$, and $E_{\rm{theory}}$ are 0.57, 0.26, and 0.17 eV with respect to experiment, with respective MAREs of 27\%, 11\%, and 6\% (see also Table \ref{final}).
The finally achieved accuracy of 0.17 eV (6\%) informs us that the 
applied corrections are essential to achieve this level of agreement with experiment,
since they are of the same size or larger on average. Furthermore, 
the achieved accuracy is comparable to that of state-of-the-art $GW$ calculations for extended sets of materials.\cite{tal2021vertex,leppert2019towards,chen2015accurate,shishkin2007accurate}

To support the use of the material-specific functional DSH, we carry out a comparison starting from standard functionals, such as PBE, HSE06, and PBE0(0.25). 
As discussed, our band-gap corrections can all accurately be calculated using the 
PBE functional. The use of a different functional then only affects the calculation
of the bare band gap. The detailed results of these calculations are given 
in the Supplementary Tables 8, 9, and 10 for PBE, HSE06, and PBE0(0.25), respectively, and can 
be directly compared with the DSH results given in Table \ref{final}.
To assess the performance of all the investigated functionals, we have summarized in 
Table \ref{MAE} their respective MAEs and MAREs for the calculated band gaps 
with respect to the experimental values.
We notice that for the PBE and HSE06 functionals the agreement with experimental band gaps steadily deteriorates upon including the corrections. 
In particular, the $E_{\rm bare}$ from HSE06 already shows relatively low MAE and MARE of 0.36 eV and 14\%, respectively. However, this result should be considered 
accidental and due to error cancellation. The results in Table \ref{MAE} also indicate that the performance of PBE0(0.25) for the materials investigated is rather good 
(MAE of 0.28 eV; MARE of 11\%) showing systematic improvement upon the inclusion 
of the corrections. 
As seen in Table \ref{MAE}, DSH outperforms all the other functionals with a MAE 
of 0.17 eV and a MARE of 6\%. We stress that the best agreement between theory and experiment is obtained when all the various contributing physical effects are properly estimated. 

In hindsight, it is of interest to question whether the approach developed is indeed superior
to simple band-gap predictions made on the basis of a trivial PBE calculations. To address this issue, we 
develop a simple model based on the PBE band gap and test it on the nine materials for which experimental 
band gaps are available. From Supplementary Fig.\ 9, one infers that the relation between PBE and experimental 
band gaps is approximately linear. To obtain a model prediction, we then fit the data with a linear function: 
$E_{\rm gap}^{\rm model}=1.39\cdot E_{\rm gap}^{\rm PBE} + 0.47$ eV. Such a model prediction yields a MAE of 
0.30 eV with a maximum error as high as 0.92 eV. These errors correspond to the data fitted and 
are susceptible to increase when a larger set of materials is taken under consideration. 
Nevertheless, the MAE value and the maximum error of this simple model are already noticeably worse than 
the MAE of 0.17 eV and maximal error of 0.49 eV found with the scheme introduced in this work. 
This provides further support to the validity of our scheme.

We remark that the present methodology has here successfully been applied to 
the large class of noncubic metal halide perovskite materials. To validate this methodology 
to this class of materials, we verified that the consideration of anharmonicity, cross-coupling between SOC and phonon effects, and nonadiabatic effects in phonon renormalization due to long-range polar interactions, all lead to negligible contributions
to the band gap. Thus, it should be understood that the present methodology can reach the 
high level of accuracy found in our work only to extent that these conditions hold.

To summarize, we have investigated the fundamental band gaps for a set
of inorganic halide perovskites through the use of various functionals. 
To ensure a meaningful comparison with experiment, we have included band-gap
corrections due to spin-orbit coupling and to thermal 
vibrations including zero-point motions. In particular, we use the dielectric-dependent 
hybrid functional DSH with parameters fixed through the dielectric function. 
Among the functionals considered, the functional DSH stands out providing a 
high level of accuracy compared to experimental band gaps ($\text{MAE} = 0.17$ eV; $\text{MARE} = 6$\%). The achieved accuracy relies on the consideration of 
both corrections, which lead to a step by step improvement of the agreement 
with experiment. Our final accuracy is comparable to the state of the art 
as far as band-gap estimates are concerned.

Additionally, our scheme is designed to be computationally efficient. 
The band gap calculations with the hybrid functional are performed only once 
for a single crystalline structure for each material. 
We demonstrate that both band-gap corrections due to spin-orbit coupling 
and to thermal vibrations including zero-point motions can accurately 
be determined with the computationally more convenient semilocal PBE functional.
In particular, the effects due to zero-point motions and 
thermal vibrations are estimated through the special displacement 
method, which gives the band-gap correction through a one-shot 
calculation, thereby speeding up the effort with respect to the more lengthy molecular 
dynamics simulations. Consequently, our scheme is not only highly accurate but
also provides band gaps in a computationally efficient way. These features make our
approach thus convenient for material screening procedures\cite{im2019identifying, yan2017solar, pilania2016machine} in which large databases of metal halide perovskites are taken under consideration. 

%


\section{Methods}\label{sec4}
\subsection{Range-separated dielectric-dependent hybrid functional}

We use a range-separated hybrid functional formulation in which the parameters are determined from the dielectric response of the system. 
In such a scheme, the Coulomb potential is partitioned into a short-range (SR) and a long-range (LR) component through an error 
function \cite{heyd2003hybrid,skone2014self,skone2016nonempirical,chen2018nonempirical,cui2018doubly,liu2019assessing}:

\begin{equation}
  \frac{1}{\vert \textbf r-\textbf r' \vert} = 
  \underbrace{\frac{{\rm erfc}(\mu \vert \textbf r-\textbf r' \vert)}{\vert \textbf r-\textbf r' \vert}}_{\text{SR}} +
  \underbrace{\frac{{\rm erf}(\mu \bm{\vert \textbf r-\textbf r' \vert)}}{\vert \textbf r-\textbf r' \vert}}_{\text{LR}} ,
  \label{range-separation}
\end{equation} 
where $\mu$ is the range-separation parameter.
Next, the SR and LR components appearing in the nonlocal exchange potential are separately decomposed into nonlocal Fock 
and semilocal PBE \cite{Perdew1996} exchange terms through admixtures defined by $\alpha_{\rm{SR}}$ and $\alpha_{\rm{LR}}$:
\begin{eqnarray}
  V_{\rm{X}}(\textbf r,\textbf r') &= 
  &\alpha_{\rm{SR}} V_{\rm{X}}^{\rm{Fock,SR}}(\textbf r,\textbf r';\mu) + (1 - \alpha_{\rm{SR}})V_{\rm{X}}^{\rm{PBE,SR}}(\textbf r; \mu)\delta(\textbf r - \textbf r) \nonumber\\
  &&+  \alpha_{\rm{LR}} V_{\rm{X}}^{\rm{Fock,LR}}(\textbf r,\textbf r';\mu) + (1 - \alpha_{\rm{LR}})V_{\rm{X}}^{\rm{PBE,LR}}(\textbf r; \mu)\delta(\textbf r-\textbf r').
  \label{ex-potential}
\end{eqnarray}
Numerous widely used density functionals can be recovered from Eq. (\ref{ex-potential}). For example, irrespective of the choice of $\mu$, the semilocal functional PBE (Ref. \citenum{Perdew1996}) is found for $\alpha_{\rm{SR}}=\alpha_{\rm{LR}}=0$, whereas the global hybrid functional PBE0 (Ref. \citenum{perdew1996rationale}) can be obtained by setting 
$\alpha_{\rm{SR}}=\alpha_{\rm{LR}}=0.25$. Similarly, the two recently proposed DDH functionals, i.e.\ DD-RSH-CAM \cite{chen2018nonempirical} and DSH \cite{cui2018doubly}, also belong to this class of functionals and can be found by setting $\alpha_{\rm{SR}} = 1$ and $\alpha_{\rm{LR}} = {1}/{\epsilon_{\infty}}$. However, these two DDH schemes differ in the way the parameter $\mu$ is set. 
In DD-RSH-CAM, $\mu$ is derived from a fit to the dielectric function calculated through 
first-principles linear response \cite{chen2018nonempirical,bischoff2019nonempirical}, 
while in DSH this parameter is approximated through an empirical expression that can be evaluated analytically \cite{cui2018doubly}. As we show below, the two proposed values of $\mu$ 
are nevertheless rather close (see Supplementary Table 4). Therefore, for speeding up 
the computational effort, we adopt in this work the dielectric-dependent functional 
denoted DSH in Ref. \citenum{cui2018doubly} with the empirical expression for $\mu$ 
proposed therein.

More specifically, the empirical expression for $\mu$ proposed in ref. \citenum{cui2018doubly} stems from the nonlocal screened Coulomb potential derived by Shimazaki and Asai \cite{shimazaki2008band}:
\begin{equation}
  V(\bm{r,r'}) = \left(1-\frac{1}{\epsilon_{\infty}}\right)\frac{{\rm exp}(-\tilde{k}_{\rm{TF}} \vert \textbf r-\textbf r'\vert)}{\vert \textbf r-\textbf r'\vert}  +  \frac{1}{\epsilon_{\infty}}\frac{1}{\vert \textbf r-\textbf r'\vert}.
  \label{ScreenedPotential-exp}
\end{equation}
Here, $\tilde{k}_{\rm{TF}}^2 = {k_{\rm{TF}}^2} \left[\left(\epsilon_\infty - 1\right)^{-1} + 1\right]/\alpha$ and $k_{\rm{TF}}=2\sqrt[6]{(3n/\uppi)}$ is the Thomas-Fermi screening length, which depends on the valence charge density \textit{n}. Following refs\ \citenum{cui2018doubly,skone2016nonempirical}, we include the \textit{d} electrons in the valence density
when the valence band is dominated by \textit{d} orbitals, namely for Ge, Sn, Pb, Br, and I. The coefficient $\alpha=1.563$ is considered to be independent of material.\cite{bechstedt1992efficient,cappellini1993model}
Equation (\ref{ScreenedPotential-exp}) can be well approximated by the error-function expression \cite{shimazaki2008band},
\begin{equation}
  V(\textbf r,\textbf r') 
  =\left(1-\frac{1}{\epsilon_{\infty}}\right)\frac{{\rm erfc}(\mu \vert \textbf r- \textbf r'\vert)}{\vert \textbf r - \textbf r'\vert}  +  \frac{1}{\epsilon_{\infty}}\frac{1}{\vert \textbf r-\textbf r'\vert},
  \label{ScreenedPotential-erf}
\end{equation}
provided one sets $\mu = {2}\tilde{k}_{\rm{TF}}/3$. 
In Eqs. (\ref{ScreenedPotential-exp}) and (\ref{ScreenedPotential-erf}), the first term considers only the SR contribution that corresponds to metallic screening, whereas the second term considers the full-range (FR) of the Coulomb interaction representing dielectric screening.\cite{cui2018doubly} Therefore, the nonlocal Fock exchange found by Shimazaki and Asai can be expressed as:\cite{cui2018doubly} 
\begin{equation}
\begin{aligned}
  V_{\rm X}^{\rm nonlocal}(\textbf r,\textbf r') &=\left(1-\frac{1}{\epsilon_{\infty}}\right)V{\rm{_{X}^{Fock,SR}}}(\textbf r,\textbf r'; \mu)   +   \frac{1}{\epsilon_{\infty}}V{\rm{_{X}^{Fock,FR}}}(\textbf r,\textbf r') \\
  & = \left(1-\frac{1}{\epsilon_{\infty}}\right)V{\rm{_{X}^{Fock,SR}}}(\textbf r,\textbf r'; \mu)  \\
  & +    \frac{1}{\epsilon_{\infty}}V{\rm{_{X}^{Fock,LR}}}(\textbf r,\textbf r'; \mu) + \frac{1}{\epsilon_{\infty}}V{\rm{_{X}^{Fock,SR}}}(\textbf r,\textbf r'; \mu)\\
  & = V{\rm{_{X}^{Fock,SR}}}(\textbf r, \textbf r';\mu)   +  \frac{1}{\epsilon_{\infty}}V{\rm{_{X}^{Fock,LR}}}(\textbf r, \textbf r';\mu).
  \label{DSH}
\end{aligned}
\end{equation}
To use this expression as nonlocal exchange in a hybrid functional formulation, one needs to add a 
compensating semilocal exchange in the long range.\cite{cui2018doubly} Thus, one obtains
\begin{eqnarray}
  V_{\rm{X}}^{\rm DSH}(\textbf r,\textbf r') &=& V_{\rm{X}}^{\rm{Fock,SR}}(\textbf r,\textbf r';\mu) + 
  \frac{1}{\epsilon_{\infty}} V_{\rm{X}}^{\rm{Fock,LR}}(\textbf r,\textbf r';\mu) \nonumber \\ 
  && + (1 - \frac{1}{\epsilon_{\infty}})V_{\rm{X}}^{\rm{PBE,LR}}(\textbf r; \mu)\delta(\textbf r-\textbf r').
  \label{ex-potential-new}
\end{eqnarray}
This expression is formally the same as the one obtained by Chen {\it et al.}\
following an alternative derivation path \cite{chen2018nonempirical}. This expression is also 
consistent with Eq. (\ref{ex-potential}), through which it can be obtained by setting 
$\alpha_{\rm{SR}} = 1$ and $\alpha_{\rm{LR}} = {1}/{\epsilon_{\infty}}$. Hence, we use in this work
the exchange potential given in Eq. (\ref{ex-potential-new}) with $\mu = {2}\tilde{k}_{\rm{TF}}/3$. 

\subsection{Special displacement method}\label{sec:SDM}

Based on the theory introduced by Williams \cite{williams1951theoretical} and Lax \cite{lax1952franck} (WL) and the
use of the harmonic approximation, the temperature-dependent band gap can be expressed as 
\cite{zacharias2016one}:
\begin{equation}
    E_{\rm{gap}}^T=\prod_\nu\int dQ_\nu
    \frac{\rm{exp}\left[-Q^2_\nu/\left(2\sigma^2_{\nu,\textit{T}}\right)\right]}{\sqrt{2\uppi}\sigma_{\nu,T}}E_{\rm gap}^Q,
\label{WL-T}
\end{equation}
where the product runs over all modes $\nu$, and $Q$ is used to indicate collectively the configuration defined by the normal coordinates ${Q_\nu}$. This expression can be readily understood as the thermal average of $E_{\rm{gap}}^Q$ 
with weights $\prod_\nu\rm{exp}\left[-Q^2_\nu/\left(2\sigma^2_{\nu,\textit{T}}\right)\right]/\sqrt{2\uppi}\sigma_{\nu,\textit{T}}$, in which $\sigma_{\nu,T}$ is a spatial Gaussian broadening given by
\begin{equation}\label{ForBOTH}
 \sigma_{\nu,T}=l_\nu\sqrt{2n_{\nu,T}+1}=\sigma_{{\rm ZPR}+T},
\end{equation}
with $l_\nu=\sqrt{ \hbar / (2M_{\rm p}\omega_\nu) }$ and 
$n_{\nu,T}=\{ \exp\left[\hbar\omega_\nu / \left(k_{\rm B}T\right)\right]-1\}^{-1}$ 
for the zero-point vibrational amplitude and the Bose-Einstein distribution, respectively. 
$M_{\rm p}$ and $\omega_\nu$ denote the mass of the proton and the frequency of the $\nu$th normal mode. 
The broadening $\sigma_{\nu,T}$ accounts for both the zero-point and thermal effects, and is denoted 
as $\sigma_{{\rm ZPR}+T}$. For $T \rightarrow 0$ and $n_{\nu,T} \rightarrow 0$, the broadening only results from the zero-point amplitude through
\begin{equation}\label{ForZPR}
\sigma_{\nu,T}=l_{\nu}=\sqrt{\hbar/\left(2M{_{\rm p}}\omega_\nu\right)}=\sigma_{\rm ZPR}.
\end{equation}

The average in Eq. (\ref{WL-T}) for a temperature $T$ can be obtained by considering a single 
distorted atomic configuration, which is created by displacing the atoms by an amount of $\Delta \tau_{k\alpha}$ 
from the equilibrium structure \cite{zacharias2016one}:
\begin{equation}\label{DeltaTau}
  \Delta \tau_{k\alpha}= {\sqrt \frac{M_{\rm p}}{M_{\rm k}}} \sum_\nu ^{3N-3} (-1)^{(\nu-1)}e_{k\alpha,\nu}\sigma_{\nu,T},
\end{equation}
where $\nu$ runs over all the non-translational modes, $\alpha$ indicates a Cartesian direction $x$, $y$ or $z$, 
$M_{\rm k}$ is the mass of the $k$th nucleus, and $e_{k\alpha,\nu}$ is the vibrational eigenmode of the $\nu$th 
normal mode. We note that depending on whether we use $\sigma_{\nu,T}$ from Eq. (\ref{ForBOTH}) or from Eq. (\ref{ForZPR}), we either describe the ZPR and thermal effects together or just the ZPR effects alone.
From the study of Zacharias and Giustino \cite{zacharias2016one}, we infer that the band gap calculated for 
the distorted structure defined by the displacements in Eq. (\ref{DeltaTau}) reproduces the thermally 
averaged band gap defined in Eq. (\ref{WL-T}) within an accuracy of about 50 meV when the adopted 
supercell contains 150 atoms or more. \cite{zacharias2016one} We consider this level of accuracy sufficient for the purpose of our work.

The special displacement method allows one to obtain band-gap renormalizations due to phonon-coupling within the context of the adiabatic approximation.\cite{zacharias2016one,zacharias2020theory} 
The problematic limit at small momentum transfers in the summation over the Brillouin zone \cite{ponce2015temperature} is effectively handled in our supercell calculations by considering larger and larger supercells.
We systematically considered supercells of varying size (containing a number atoms ranging between 160 to 625) and obtained constant results within 0.05 eV. In practice, the frequencies remain always finite and the $\omega_0$ of the recipe of Zacharias and Giustino \cite{zacharias2020theory} could be taken to correspond to the lowest frequency found for the series of considered supercell sizes. For any lower $\omega_0$, the results would remain unchanged and the obtained values correspond to a finite result.
However, the adiabatic approximation does not account for the long-range polar coupling to 
the longitudinal optical phonons, which results from a nonadiabatic treatment.\cite{nery2016influence} In this work, this effect has been estimated 
through the one-band Fr\"ohlich model (see Sec. 2.3 and Supplementary Note 1).


\subsection{Calculation details}
The DSH band gaps of the ideal crystalline systems are determined through the 
implementation of DD-RSH-CAM functionals \cite{chen2018nonempirical} 
available in the {\sc Q}uantum-{\sc espresso} software suites.\cite{giannozzi2009quantum}
We use fully-relativistic pseudopotentials generated by the optimized norm-conserving Vanderbilt pseudopotential scheme \cite{hamann2013optimized} to account for spin-orbit coupling. 
All the calculations are carried out with the stringent set of the Pseudo Dojo (vailable at http://www.pseudodojo.org) to ensure band gaps converged within 0.1 eV. A kinetic plane-wave cutoff of 100 Ry is used for all materials together with sufficiently dense \textbf k-point grids. 
In the band-gap calculations, we use unit cells with atomic structures and lattice parameters taken from experiment (cf.\ Supplementary Table 2).
The high-frequency dielectric constant $\epsilon_\infty$ is calculated at the PBE level through the application 
of finite electric fields \cite{umari2002ab,souza2002first} and is used to fix the parameter $\alpha_\text{LR}$ of the DSH functional.

For the application of the special displacement method,\cite{zacharias2016one,zacharias2020theory} we use the {\sc cp2k} suite of codes, \cite{hutter2014cp2k} which comprises an efficient tool for determining the vibrational frequencies and modes through finite differences. The core-valence interactions are described through Goedecker-Teter-Hutter (GTH) pseudopotentials.\cite{goedecker1996separable, hartwigsen1998relativistic} We use double-zeta basis sets of MOLOPT quality.\cite{vandevondele2007gaussian} The plane-wave energy cutoff for the electron density is set to 600 Ry to ensure the convergence of the total energy. 
The vibrational frequencies $\omega_\nu$ and eigenmodes $e_{k\alpha,\nu}$ required 
for the distorted structure specified in Eq. (\ref{DeltaTau}) are obtained 
through finite displacements of 0.01 \AA\ from fully relaxed atomic positions. 
We use $\Gamma$-point sampling with the supercells defined in the Supplementary Table 5.
As an alternative scheme for obtaining vibrational properties, the non-diagonal approach 
proposed by Lloyd-Williams et al.\ \cite{lloyd2015lattice} could be used to further 
reduce the involved computational cost.

The band gaps are determined from linear extrapolations of the EDOS obtained for the 
supercells specified in the Supplementary Table 5. The {\sc cp2k} results are 
systematically obtained at the PBE level of theory, with the exception of the test results
in Table \ref{phonon-functional}, which required the use of hybrid functionals. 
In this case, the auxiliary density matrix method (ADMM) is employed to speed up 
the calculations.\cite{guidon2010auxiliary}

Our efficient band-gap calculations contain several parts that contribute to the overall 
computational cost. The first contribution results from the determination of the dielectric constant, but the cost of this calculation is not significant since it is carried out at the PBE level.
The first important contribution comes from the band-gap calculation for
the ideal crystalline system with the DSH functional. Since the cost of this calculation
is easy to evaluate, we use it here to set the unit. The calculation of the SOC correction is
carried out at the PBE level and is negligible on the scale of our cost unit.
The second important contribution comes from the determination of the vibrational properties 
at the PBE level. The corresponding cost of this part depends on the number atoms considered, but it is comparable to the cost unit for systems containing 270 atoms. 
For some materials, the EDOS needs to be determined with rather large supercells, 
but still at the PBE level. For the largest case considered (2880 atoms), this part 
gives a cost that amounts to half a cost unit and is thus quite negligible in most cases. 
All together, the dominant costs for the band-gap determination result from 
the vibrational properties and from the DSH hybrid-functional calculation 
in comparable amounts.

\backmatter

\bmhead{Data availability statement}
The structure used in the paper can be found in the materials cloud: https://archive.materialscloud.org/record/2022.35.

\bmhead{Code availability statement}
The relevant codes in this study are available from the corresponding authors upon the reasonable request.

\bmhead{Acknowledgments}

The authors acknowledge useful interactions with W. Chen, S. Falletta, M. Zacharias, and R. Ouyang. Support from the Swiss National
Science Foundation (SNSF) is acknowledged under Grant No.\ 200020-172524. The calculations have been performed at the Swiss National 
Supercomputing Center (CSCS) (grant under project ID s879 and s1123) and 
at SCITAS-EPFL.

\bmhead{Author contributions}
A.P. and H.W. designed this project. H.W. performed most calculations with the help from T.B. and P.G. for some materials. A.T. carried out calculations using VASP. H.W. and A.P. wrote the manuscript. All authors discussed data and reviewed the manuscript.

\bmhead{Competing interests}
The authors declare no competing interests.

\bibliography{sn-bibliography}

\bmhead{Figure captions}
\bmhead{Figure 1: Band gap accuracy with respect to experiment} Mean absolute errors (MAEs) of band gaps calculated with various functionals with respect to the QS$GW$ values \cite{huang2016electronic,mckechnie2018dynamic}. The percentages correspond to the mean absolute relative errors (MAREs).

\bmhead{Figure 2: Size of band gap corrections} Partial band-gap corrections \textbf a $\Delta E_{\rm{SOC}}$ and \textbf b $\Delta E_{{\rm ZPR}+T}$, 
    together with \textbf c the full band gap correction $\Delta E_{\rm{full}}=\Delta E_{\rm{SOC}} + \Delta E_{{\rm ZPR}+T}$ for all the materials in this work. \label{figure2}.

\bmhead{Figure 3: Band gap determination through linear extrapolations} \textbf a EDOS calculated within PBE with {\sc vasp} for a primitive cell of $\gamma$-\ce{CsSnI3} using a \textbf k-point mesh of 16$\times$16$\times$16 and a Gaussian smearing of $\sigma=0.05$ eV, compared with a 6$\times$4$\times$6 supercell calculation with {\sc cp2k} 
    using the $\Gamma$ point and $\sigma=0.15$ eV. \textbf{b} EDOS for the ideal structure and for the displaced structures accounting for ZPR and ZPR+$T$ ($T = 300$ K) in the case of M-\ce{RbSnF3} ($\sigma=0.15$, PBE). \textbf{c} and \textbf{d} show the valence and conduction edges, respectively. Dashed lines indicate the linear extrapolations used to determine the VBM and CBM. \label{figure3} The extracted band edges are given for the cases defined by the color code in \textbf{b}.

\bmhead{Figure 4: Calculated vs.\ measured band gaps} Band gaps obtained with the DSH functional $E_{\rm bare}$ (yellow triangles), after SOC correction $E_{\rm{SOC}}$ (purple circles), and additionally including ZPR and thermal effects $E_{\rm theory}$ (green hexagons). The MAEs and MAREs are given with respect to the experimental values (red squares).

\newpage
\begin{table}[htb]
  \caption{{\bf Parameters defining the DSH functional} Dielectric constants $\epsilon_{\infty}$ calculated at the PBE level compared to previous results from Ref. \citenum{petousis2017high}, the corresponding fraction of Fock exchange in the long range $\alpha_{\rm{LR}}=1/\epsilon_{\infty}$, and the inverse screening length $\mu$.}
  \label{DielectricConstant}  
  \centering
  \begin{tabular}{rccccc}
    \hline
     &  \multicolumn{2}{c}{$\epsilon_{\infty}$} & \multicolumn{2}{c}{$\alpha_{\rm{LR}}$}& $\mu$\\ 
    \cmidrule(r){2-3} \cmidrule(r){4-5}
              &  current work& benchmarks & current work& benchmarks & bohr$^{-1}$ \\
    \hline
    R-\ce{CsGeCl3}& 3.60&  3.64$^\textit{a}$ & 0.28&  0.27$^\textit{a}$&   0.71\\
    R-\ce{CsGeBr3}& 4.76&  & 0.21&   & 0.73\\ 
    R-\ce{CsGeI3} & 6.33&  6.13$^\textit{a}$&  0.16&  0.16$^\textit{a}$&  0.69\\ 
    R-\ce{CsPbF3} & 3.08&  & 0.32&  &  0.78\\
    \hline
    $\gamma$-\ce{RbGeBr3}& 4.15&  3.98$^\textit{a}$&  0.24&  0.25$^\textit{a}$&  0.75\\ 
    $\gamma$-\ce{RbGeI3} & 5.10&  5.09$^\textit{b}$&  0.20&  0.20$^\textit{b}$&  0.71\\
    $\gamma$-\ce{CsSnBr3}& 5.73&  &  0.17&  &  0.71\\ 
    $\gamma$-\ce{CsSnI3} & 7.24&  &  0.14&  &  0.67\\
    $\gamma$-\ce{CsPbCl3}& 3.77&  &  0.27&  &  0.69\\
    $\gamma$-\ce{CsPbBr3}& 4.58&  4.21$^\textit{a}$&  0.22&  0.24$^\textit{a}$&  0.73\\ 
    $\gamma$-\ce{CsPbI3} & 5.45&  &  0.18&  &  0.69\\
    \hline
    $\delta$-\ce{RbGeI3} & 5.16&  &  0.19&  &  0.72\\
    $\delta$-\ce{RbSnI3} & 5.01&  4.84$^\textit{a}$&  0.20&  0.21$^\textit{a}$&  0.71\\ 
    $\delta$-\ce{RbPbI3} & 4.78&  4.54$^\textit{a}$&  0.21&  0.22$^\textit{a}$&  0.71\\
    $\delta$-\ce{CsSnI3} & 4.87&  4.71$^\textit{a}$&  0.21&  0.21$^\textit{a}$&  0.71\\
    $\delta$-\ce{CsPbBr3}& 3.89&  &  0.26&  &  0.75\\
    $\delta$-\ce{CsPbI3} & 4.67&  4.43$^\textit{a}$&  0.21&  0.23$^\textit{a}$&  0.71\\ 
    \hline
    $\beta$-\ce{CsSnI3} & 7.20&   8.04$^\textit{a}$&  0.14&  0.12$^\textit{a}$&  0.67\\
    \hline
    M-\ce{RbSnF3} & 2.62&   &        0.38&  &  0.82\\
    M-\ce{CsSnF3} & 2.65&   &        0.38&  &  0.81\\
    M-\ce{CsSnCl3}& 3.38&  3.15$^\textit{b}$&  0.30&  0.32$^\textit{b}$&  0.71\\ 
    \hline
  \end{tabular}
  $^\textit{a}$~Ref.~\citenum{petousis2017high}.
  $^\textit{b}$ Obtained in this work with the DSH functional.
\end{table}

\newpage
\begin{table}[t!]
  \caption{{\bf Band gaps as calculated with various schemes} Fundamental band gaps (in eV), $E_{\rm{bare}} = E_{\rm{CBM}} - E_{\rm{VBM}}$, as calculated with the functionals PBE, HSE06, PBE0(0.25), and DSH. The VBM and CBM are the valence band maximum and conduction band minimum obtained from Kohn-Sham energy levels. The QS$GW$ band gaps are taken from Refs.\ \citenum{huang2016electronic, mckechnie2018dynamic}.
  The MAE and MARE are given with respect to the QS$GW$ values. The last column describes the nature of the band gap, direct (D) or indirect (I).}
  \centering
  \label{bare} 
  \begin{tabular}{rcccccc}
    \hline
    & \multicolumn{5}{c}{$E_{\rm bare}$} \\
    \cmidrule(r){2-6}
    &  PBE & HSE06 & PBE0(0.25) &   DSH & QS$GW$ & D/I\\
    \hline
    R-\ce{CsGeCl3}& 2.11&  2.95&  3.71&   4.01&  4.37$^\textit{a}$&  D\\ 
    R-\ce{CsGeBr3}& 1.21&  2.10&  2.85&   3.05&  2.70$^\textit{a}$&  D\\ 
    R-\ce{CsGeI3} & 0.83&  1.55&  2.29&   2.17&  1.69$^\textit{a}$&  D\\ 
    R-\ce{CsPbF3} & 3.31&  4.33&  5.04&   5.77& &D  \\
    \hline
    $\gamma$-\ce{RbGeBr3}& 1.82&  2.68&   3.40&   3.75& &  D\\  
    $\gamma$-\ce{RbGeI3} & 1.44&  2.18&   2.87&   2.94& & D\\ 
    $\gamma$-\ce{CsSnBr3}& 0.75&  1.37&   2.03&   2.03& & D\\
    $\gamma$-\ce{CsSnI3} & 0.62&  1.12&   1.75&   1.55&  1.50$^\textit{a}$&  D\\ 
    $\gamma$-\ce{CsPbCl3}& 2.05&  2.74&   3.41&   3.57& & D\\   
    $\gamma$-\ce{CsPbBr3}& 1.69&  2.43&   3.09&   3.36& & D\\
    $\gamma$-\ce{CsPbI3} & 1.56&  2.19&   2.83&   2.87 &  2.81$^\textit{b}$& D\\ 
    \hline
    $\delta$-\ce{RbGeI3} & 2.21&  3.06&   3.74&   3.74& & I\\   
    $\delta$-\ce{RbSnI3} & 2.00&  2.81&   3.48&   3.51& & I\\ 
    $\delta$-\ce{RbPbI3} & 2.39&  3.30&   3.97&   4.10& & I\\ 
    $\delta$-\ce{CsSnI3} & 1.97&  2.76&   3.43&   3.44& & I\\
    $\delta$-\ce{CsPbBr3}& 2.79&  3.84&   4.53&   5.00& & I\\     
    $\delta$-\ce{CsPbI3} & 2.44&  3.31&   3.99&   4.14& & I\\
    \hline
    $\beta$-\ce{CsSnI3} &0.45&  1.03&   1.66&   1.48&  1.49$^\textit{a}$& D\\
    \hline
    M-\ce{RbSnF3} & 3.66&  4.63&  5.36&   6.25& & D\\
    M-\ce{CsSnF3} & 3.79&  4.88&  5.61&   6.59& & I\\
    M-\ce{CsSnCl3}& 2.64&  3.43&  4.14&   4.50& &  D\\
    \hline
    MAE&           1.30&   0.64&  0.31&     0.22 &  \\
    MARE&       55\%&     24\%&    14\%&       9\%& \\
    \hline
  \end{tabular}
  
$^\textit{a}$ Ref. \citenum{huang2016electronic}; $^\textit{b}$ Ref. \citenum{mckechnie2018dynamic}.
\end{table}

\newpage
\begin{table}[t]
  \caption{{\bf Band gap corrections due to SOC} The corrections are evaluated with three different functionals: PBE, PBE0(0.25), and DSH. We give $\Delta E_{\rm{SOC}} = E_{\rm{SOC}} - E_{\rm{bare}}$, where $E_{\rm{SOC}}$ is the fundamental band gap including SOC effects.
  The band gaps are given in eV.}
  \label{SOC} 
  \centering
  \begin{tabular}{rccc}
    \hline
    &  \multicolumn{3}{c}{$\Delta E_{\rm{SOC}}$} \\
    \cmidrule(r){2-4} 
    & PBE & PBE0(0.25) & DSH \\
    \hline
    R-\ce{CsGeCl3}& $-$0.07& $-$0.07& $-$0.06  \\ 
    R-\ce{CsGeBr3}& $-$0.05& $-$0.06& $-$0.07  \\ 
    R-\ce{CsGeI3} & $-$0.14& $-$0.15& $-$0.16  \\ 
    R-\ce{CsPbF3} & $-$0.88& $-$0.85& $-$0.78  \\
    \hline
    $\gamma$-\ce{RbGeBr3}& $-$0.06& $-$0.11& $-$0.12 \\  
    $\gamma$-\ce{RbGeI3} & $-$0.11& $-$0.14& $-$0.15 \\ 
    $\gamma$-\ce{CsSnBr3}& $-$0.30& $-$0.31& $-$0.30 \\
    $\gamma$-\ce{CsSnI3} & $-$0.39& $-$0.42& $-$0.41 \\ 
    $\gamma$-\ce{CsPbCl3}& $-$1.04& $-$1.08& $-$1.03 \\   
    $\gamma$-\ce{CsPbBr3}& $-$1.02& $-$1.09& $-$1.06 \\
    $\gamma$-\ce{CsPbI3} & $-$0.99& $-$1.10& $-$1.08 \\ 
    \hline
    $\delta$-\ce{RbGeI3} & $-$0.19& $-$0.20& $-$0.20 \\   
    $\delta$-\ce{RbSnI3} & $-$0.17& $-$0.18& $-$0.18 \\ 
    $\delta$-\ce{RbPbI3} & $-$0.55& $-$0.63& $-$0.61 \\ 
    $\delta$-\ce{CsSnI3} & $-$0.17& $-$0.18& $-$0.18 \\
    $\delta$-\ce{CsPbBr3}& $-$0.37& $-$0.38& $-$0.37 \\     
    $\delta$-\ce{CsPbI3} & $-$0.55& $-$0.61& $-$0.59 \\
    \hline
    $\beta$-\ce{CsSnI3} & $-$0.37& $-$0.45& $-$0.43  \\
    \hline
    M-\ce{RbSnF3} & $-$0.01& $-$0.01& $-$0.01  \\
    M-\ce{CsSnF3} & $-$0.02& $-$0.01& $-$0.01  \\
    M-\ce{CsSnCl3}& $-$0.04& $-$0.04& $-$0.03 \\
    \hline
  \end{tabular}
\end{table}

\newpage
\begin{table}[t!]
  \caption{{\bf Calculated band gaps based on DSH} Calculated band gaps (in eV) $E_{\rm theory}=E_{\rm bare}+\Delta E{_{\rm SOC}}+\Delta E_{{\rm ZPR}+T}$, where $E_{\rm bare}$ is obtained through the DSH functional (Table \ref{bare}), $\Delta E{_{\rm SOC}}$ is the correction due to SOC (Table \ref{SOC}), and $\Delta E_{{\rm ZPR}+T}$ the correction accounting for ZPR and thermal vibrations (Table \ref{thermal+ZPR}).
  The MAEs and MAREs are given with respect to the band gaps $E_{\rm Expt.}$ measured in optical experiments.}
  \label{final} 
  \centering
  \begin{tabular}{rrrrrrc}
    \hline
     & $E_{\rm bare}$& $\Delta E_{\rm SOC}$& $E_{\rm SOC}$& $\Delta E_{{\rm ZPR}+T}$& $E_{\rm theory}$& $E_{\rm Expt.}$\\
    \hline
    R-\ce{CsGeCl3}& 4.01&  $-$0.07& 3.94&  $-$0.40& 3.54& 3.43$^\textit{a}$\\
    R-\ce{CsGeBr3}& 3.05&  $-$0.05& 3.00&  $-$0.32& 2.68& 2.38$^\textit{a}$\\
    R-\ce{CsGeI3} & 2.17&  $-$0.14& 2.03&  $-$0.26& 1.77& 1.63$^\textit{b}$\\ 
    R-\ce{CsPbF3} & 5.77&  $-$0.88& 4.89&  $-$0.10& 4.79&   \\
    \hline
    $\gamma$-\ce{RbGeBr3}& 3.75&  $-$0.06&  3.69&  $-$0.10& 3.59& \\
    $\gamma$-\ce{RbGeI3} & 2.94&  $-$0.11&  2.83&  $-$0.26& 2.57& \\
    $\gamma$-\ce{CsSnBr3}& 2.03&  $-$0.30&  1.73&   0.11& 1.84& 1.81$^\textit{c}$\\
    $\gamma$-\ce{CsSnI3} & 1.55&  $-$0.39&  1.16&   0.11& 1.27& 1.25$^\textit{c}$\\
    $\gamma$-\ce{CsPbCl3}& 3.57&  $-$1.04& 2.53&  $-$0.03& 2.50& 2.99$^\textit{c}$\\
    $\gamma$-\ce{CsPbBr3}& 3.36&  $-$1.02& 2.34&  0.00& 2.34& 2.31$^\textit{c}$\\
    $\gamma$-\ce{CsPbI3} & 2.87&  $-$0.99& 1.88&  $-$0.07& 1.81& 1.72$^\textit{c}$\\
    \hline
    $\delta$-\ce{RbGeI3} & 3.74& $-$0.19& 3.55&   $-$0.06& 3.49& \\    
    $\delta$-\ce{RbSnI3} & 3.51& $-$0.17& 3.34&   $-$0.10& 3.24& \\
    $\delta$-\ce{RbPbI3} & 4.10& $-$0.55& 3.55&   $-$0.25& 3.30& \\
    $\delta$-\ce{CsSnI3} & 3.44& $-$0.17& 3.27&   $-$0.24& 3.03& \\
    $\delta$-\ce{CsPbBr3}& 5.00& $-$0.37& 4.63&   $-$0.33& 4.30& \\    
    $\delta$-\ce{CsPbI3} & 4.14& $-$0.55& 3.59&   $-$0.19& 3.40& \\
    \hline
    $\beta$-\ce{CsSnI3} & 1.48& $-$0.37& 1.11&   0.23& 1.34& \\
    \hline
    M-\ce{RbSnF3} & 6.25& $-$0.01& 6.24&  $-$0.28& 5.96& \\
    M-\ce{CsSnF3} & 6.59& $-$0.02& 6.57&  $-$0.26& 6.31& \\
    M-\ce{CsSnCl3}& 4.50& $-$0.04& 4.46&  $-$0.34& 4.12& 4.46$^\textit{d,e}$\\
    \hline
    MAE &    0.57& & 0.26&& 0.17& \\
    MARE&  27\%&  & 11\%& &6\%& \\
    \hline
  \end{tabular}
  
  $^\textit{a}$ Ref. \citenum{lin2008study}, $^\textit{b}$ Ref. \citenum{krishnamoorthy2015lead}, $^\textit{c}$ Ref. \citenum{tao2019absolute}, $^\textit{d}$ Ref. \citenum{huang2013electronic}, and $^\textit{e}$ Ref. \citenum{voloshinovskii1994luminescence}.
\end{table}

\newpage
\begin{table}[t]
  \caption{{\bf Dependence of thermal band gap correction on functional} Band gap correction (in eV) $\Delta E_{{\rm ZPR}+T}$ due to ZPR+$T$ for M-\ce{RbSnF3}, M-\ce{CsSnF3}, and R-\ce{CsGeCl3} calculated using different functionals: PBE, PBE0(0.25), and PBE0(0.40). For more details refer to Supplementary Fig. 8.}
 \label{phonon-functional} 
  \centering
  \begin{tabular}{rccc}
    \hline
   materials & \multicolumn{3}{c}{$\Delta E_{{\rm ZPR}+T}$} \\
    \cmidrule(r){2-4}
    & PBE & PBE0(0.25) & PBE0(0.40) \\
    \hline
    M-\ce{RbSnF3} & $-$0.28& $-$0.31& $-$0.30 \\
    M-\ce{CsSnF3} & $-$0.26& $-$0.28& $-$0.28 \\
    R-\ce{CsGeCl3}& $-$0.40& $-$0.39& $-$\\
    \hline
  \end{tabular}
\end{table}

\begin{table}[t]
  \caption{{\bf Thermal band gap corrections} Band gap renormalization (in eV) $\Delta E_{{\rm ZPR}+T}$ (in eV) due to the combined effect of zero-point motion and thermal vibrations, as obtained from linear extrapolations of the wings in the EDOS.  $\Delta E_{\rm ZPR}$ corresponds to the renormalization at $T=0$ K and $\Delta E_{T} =  \Delta E_{{\rm ZPR}+T}-\Delta E_{\rm ZPR}$. The band gap corrections are evaluated at temperature $T$ (cf.\ Supplementary Table 2).}
  \label{thermal+ZPR} 
  \centering
  \begin{tabular}{rrrrr}
    \hline
    & $T$ (K) & $\Delta E_{\rm{ZPR}}$& $\Delta E_{\it{T}}$& $\Delta E_{\rm{ZPR}+\it{T}}$\\
    \hline
    R-\ce{CsGeCl3}& 300& $-$0.14& $-$0.26& $-$0.40\\ 
    R-\ce{CsGeBr3}& 300& $-$0.10& $-$0.22& $-$0.32\\ 
    R-\ce{CsGeI3}&  300& $-$0.08& $-$0.18& $-$0.26\\
    R-\ce{CsPbF3}&  148& $-$0.09& $-$0.01& $-$0.10\\
    \hline
    $\gamma$-\ce{RbGeBr3}&  300& $-$0.03& $-$0.07& $-$0.10\\  
    $\gamma$-\ce{RbGeI3} &  473& $-$0.03& $-$0.23& $-$0.26\\ 
    $\gamma$-\ce{CsSnBr3}&  100& 0.00& 0.11& 0.11\\
    $\gamma$-\ce{CsSnI3} &  300& 0.05& 0.06& 0.11\\ 
    $\gamma$-\ce{CsPbCl3}&  300& $-$0.03& 0.00& $-$0.03\\   
    $\gamma$-\ce{CsPbBr3}&  300& $-$0.02& 0.02& 0.00\\
    $\gamma$-\ce{CsPbI3} &  300& $-$0.01& $-$0.06& $-$0.07\\ 
    \hline
    $\delta$-\ce{RbGeI3} &  300& 0.00& $-$0.06& $-$0.06\\   
    $\delta$-\ce{RbSnI3} &  300& $-$0.01& $-$0.09& $-$0.10\\ 
    $\delta$-\ce{RbPbI3} &  300& $-$0.04& $-$0.21& $-$0.25\\ 
    $\delta$-\ce{CsSnI3} &  300& $-$0.05& $-$0.19& $-$0.24\\
    $\delta$-\ce{CsPbBr3}&  300& $-$0.06& $-$0.27& $-$0.33\\     
    $\delta$-\ce{CsPbI3} &  300& $-$0.02& $-$0.17& $-$0.19\\
    \hline
    $\beta$-\ce{CsSnI3}& 380& 0.00& 0.23& 0.23\\
    \hline
    M-\ce{RbSnF3} &  300& $-$0.20& $-$0.08& $-$0.28\\
    M-\ce{CsSnF3} &  300& $-$0.16& $-$0.10& $-$0.26\\
    M-\ce{CsSnCl3}&  300& $-$0.12& $-$0.22& $-$0.34\\
    \hline
  \end{tabular}
\end{table}

\begin{table}[t!]
  \caption{ {\bf Band gap accuracy with respect to experiment} MAEs (in eV) for band gaps obtained with various different functionals compared to experiment. The MAREs is given in parentheses.}
  \label{MAE} 
  \centering
  \begin{tabular}{lccc}
    \hline
     & $E_{\rm bare}$& $E{_{\rm SOC}}$&  $E_{\rm theory}$\\
    \hline
    PBE& 0.95 (39\%)&  1.40 (61\%) &  1.53 (66\%) \\
    HSE06& 0.36 (14\%) &  0.68 (29\%) &  0.82 (33\%) \\
    PBE0(0.25)& 0.53 (27\%) &  0.31 (13\%)&  0.28 (11\%) \\
    DSH& 0.57 (27\%)&  0.26 (11\%)&  0.17 (6\%) \\
    \hline
  \end{tabular}
\end{table}

\clearpage

\begin{figure}
    \centering
    \includegraphics[width=0.65\textwidth]{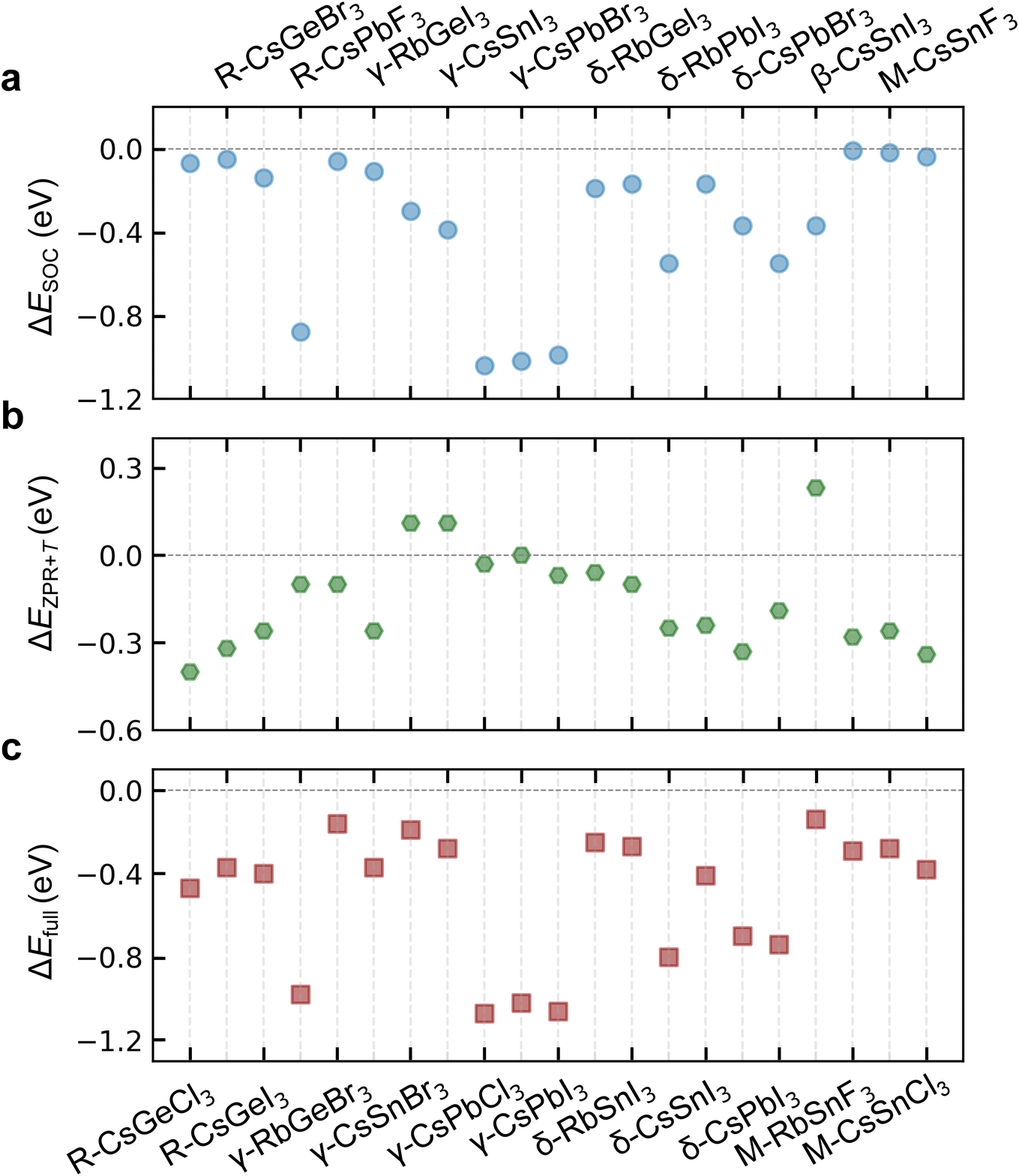}
    \caption{\textbf{Size of band gap corrections} Partial band-gap corrections \textbf a $\Delta E_{\rm{SOC}}$ and \textbf b $\Delta E_{{\rm ZPR}+T}$, 
    together with \textbf c the full band gap correction $\Delta E_{\rm{full}}=\Delta E_{\rm{SOC}} + \Delta E_{{\rm ZPR}+T}$ for all the materials in this work. \label{figure2}}
\end{figure}

\begin{figure}[htb]
    \centering
    \includegraphics[width=0.75\textwidth]{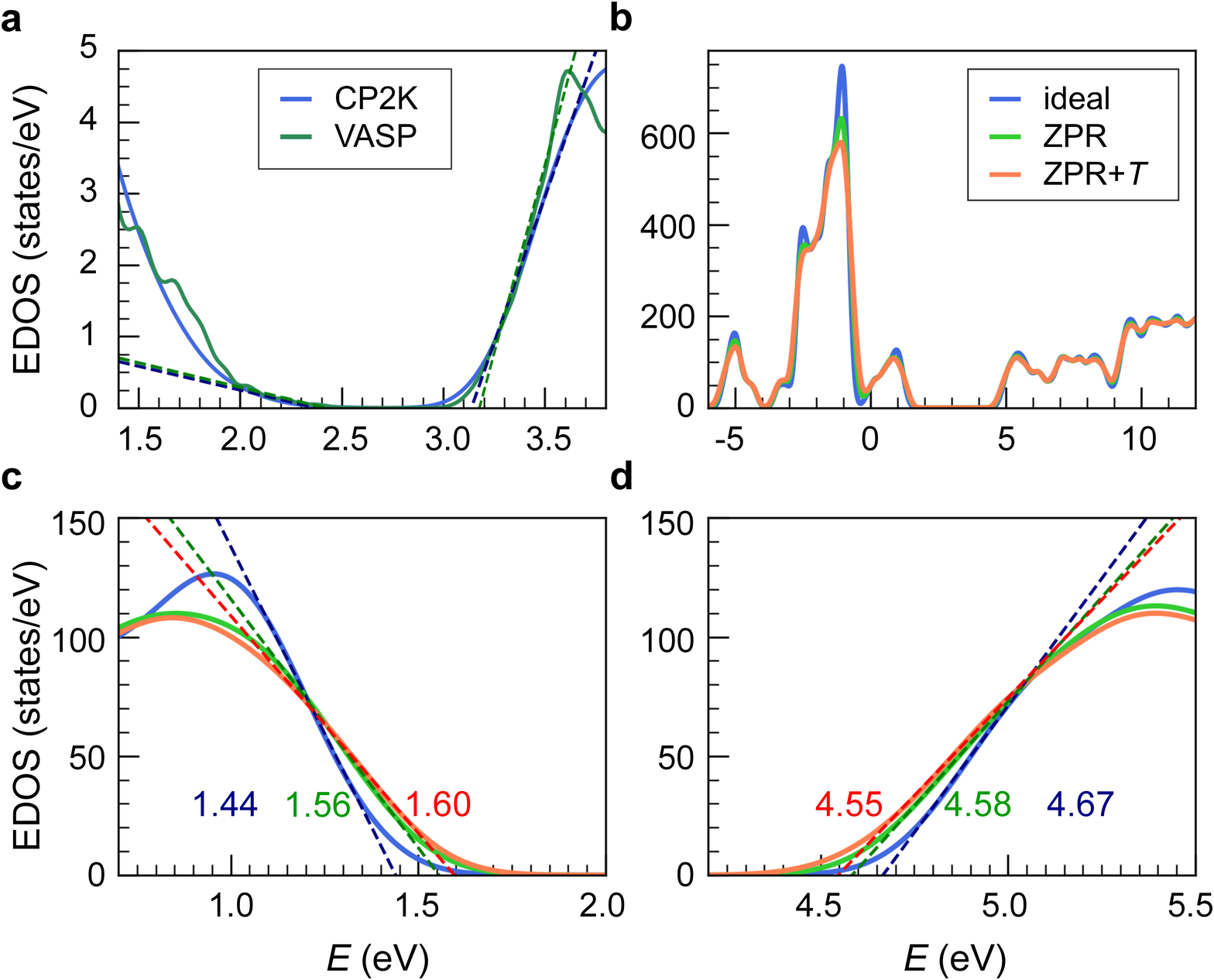}
    \caption{\textbf{Band gap determination through linear extrapolations} \textbf a EDOS calculated within PBE with {\sc vasp} for a primitive cell of $\gamma$-\ce{CsSnI3} using a \textbf k-point mesh of 16$\times$16$\times$16 and a Gaussian smearing of $\sigma=0.05$ eV, compared with a 6$\times$4$\times$6 supercell calculation with {\sc cp2k} 
    using the $\Gamma$ point and $\sigma=0.15$ eV. \textbf{b} EDOS for the ideal structure and for the displaced structures accounting for ZPR and ZPR+$T$ ($T = 300$ K) in the case of M-\ce{RbSnF3} ($\sigma=0.15$, PBE). \textbf{c} and \textbf{d} show the valence and conduction edges, respectively. Dashed lines indicate the linear extrapolations used to determine the VBM and CBM. \label{figure3} The extracted band edges are given for the cases defined by the color code in \textbf{b}.}
\end{figure}

\begin{figure}[t]
    \centering
    \includegraphics[width=0.6\textwidth]{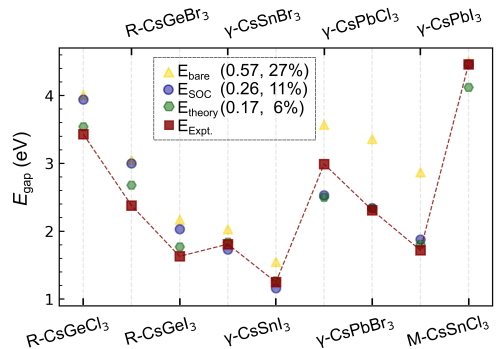}
    \caption{\textbf{Calculated vs.\ measured band gaps} Band gaps obtained with the DSH functional $E_{\rm bare}$ (yellow triangles), after SOC correction $E_{\rm{SOC}}$ (purple circles), and additionally including ZPR and thermal effects $E_{\rm theory}$ (green hexagons). The MAEs and MAREs are given with respect to the experimental values (red squares). \label{figure4}}
\end{figure}

\end{document}